\documentclass[prb,aps,twocolumn,nofootinbib,superscriptaddress]{revtex4-2}
\usepackage[usenames,dvipsnames]{color}
\usepackage[colorlinks=true,linkcolor=Blue,citecolor=Blue,urlcolor=Blue]{hyperref}
\usepackage{comment,graphicx,physics,booktabs}
\usepackage{bbm}
\usepackage{bm}
\usepackage{amsmath,bbm}
\usepackage{amssymb}
\usepackage{mathptmx}
\usepackage[version=3]{mhchem}

\renewcommand\vec{\textbf}
\def\r{{\vec{r}}}

\def\v{{\vec{v}}}

\begin{document}
\title{Reconstructing the Hamiltonian from the local density of states using neural networks}
\author{Nisarga Paul}
\affiliation{Department of Physics, Massachusetts Institute of Technology, Cambridge, Massachusetts 02139, USA}
\author{Andrew Ma}
\affiliation{Department of Electrical Engineering and Computer Science, Massachusetts Institute of Technology, Cambridge, Massachusetts 02139, USA}
\author{Kevin P. Nuckolls}
\affiliation{Department of Physics, Massachusetts Institute of Technology, Cambridge, Massachusetts 02139, USA}
\begin{abstract}
Reconstructing a quantum system's Hamiltonian from limited yet experimentally observable information is interesting both as a practical task and from a fundamental standpoint. We pose and investigate the inverse problem of reconstructing a Hamiltonian from a spatial map of the local density of states (LDOS) near a fixed energy. We demonstrate high-quality recovery of Hamiltonians from the LDOS using supervised learning. In particular, we generate synthetic data from single-particle Hamiltonians in 1D and 2D, train convolutional neural networks, and obtain models that solve the inverse problem with remarkably high accuracy. Moreover, we are able to generalize beyond the training distribution and develop models with strong robustness to noise. Finally, we comment on possible experimental applications to scanning tunneling microscopy, where we propose that maps of the electronic local density of states might be used to reveal a sample's unknown underlying energy landscape. 
\end{abstract}
\maketitle

\section{Introduction}
In this paper, we pose and study a question at the intersection of quantum condensed matter physics and inverse problems: recovering a Hamiltonian from the local density of states. Given a Hamiltonian $H$ with eigenstates $\psi_E(\vec r)$ where $\vec r$ is position, it is straightforward to compute the local density of states $\rho_E(\vec r) = \sum_{|E'-E|<\delta E} |\psi_{E'}(\vec r)|^2$ near some energy $E$. Here we address the question: can we proceed in reverse?  \par 
\paragraph*{Motivation.} This problem falls into the category of Hamiltonian learning or reconstruction problems~\cite{Bairey2019Jan,Qi2019Jul,Cao2020Nov,Hou2020Aug,Cao2020May,Gebhart2023Mar, Nandy2024May,Xia2025Jan}. Hamiltonian learning has been studied from a variety of perspectives. For example, previous studies have demonstrated analytical~\cite{Qi2019Jul,Garcia-Pintos2024Jul} and approximate numerical~\cite{Lantz2021May,Zhao2023May} Hamiltonian reconstruction from eigenstate data or measurements, and have performed Hamiltonian learning from the Gibbs state and real-time dynamics~\cite{Huang2023May,Yu2023Jun,Gu2024Jan,Haah2024Jun,Dutkiewicz2024Nov}. 
Many previous works assume the data of an entire eigenstate or density matrix is available. However, the entire quantum state, including all complex amplitudes, is typically not readily accessible.\par 
In contrast, quantities like the local density of states (LDOS) are of direct physical relevance, especially in a condensed matter context. For example, the LDOS can be measured on conducting surfaces using scanning tunneling miscroscopy (STM), an atomic-scale resolution imaging technique~\cite{Binnig1982Jul,Repp2005Jan,Bian2021May}. STM has become an indispensable tool in condensed matter physics, with modern equipment capable of producing large numbers of ultra-high-resolution images of the local electronic structure of materials~\cite{chen2021introduction,Yin2021Apr,Nuckolls2024Jul}. As a result, an emerging challenge is to extract interesting quantitative features from these images that are not readily discernible by eye. Learning the system's Hamiltonian in principle captures \textit{all} essential features, which can be computed at will in downstream tasks. 

\par 

Our work complements recent efforts that have demonstrated the utility of machine learning for extracting rich information from LDOS data, such as detecting nematic order \cite{Goetz2020Jun,Sobral2023Aug}, denoising images through self-supervised methods \cite{Kuijf2024Sep}, reconstructing effective Hamiltonians in disordered quantum materials \cite{Basak2023May}, and performing automated structure discovery for molecules \cite{Kurki2023Dec}. The utility of machine learning for Hamiltonian learning from experimental image data has also been broadly recognized~\cite{Basak2023May,Percebois2021Aug,Percebois2023Dec}. \par 

\paragraph*{Problem statement.} We focus on a simple, concrete version of the problem: a single-particle Hamiltonian of the form $H = T+V(\vec r)$ with kinetic operator $T$ and potential $V(\vec r)$ defined on a lattice. In particular, we choose a tight-binding model with lattice constant $a$ and nearest-neighbor hopping $t$ of the form
\begin{equation}
    H = -\frac{t}{2}\sum_{\langle \vec r\vec r'\rangle} c_{\vec r}^\dagger c_{\vec r'} + \sum_{\vec r} V(\vec r) c_{\vec r}^\dagger c_{\vec r},
\end{equation}
where $\langle \vec r\vec r'\rangle$ denotes a sum over nearest-neighbor pairs. This is a simple starting point to address the basic problem of this work. Denote energies and eigenstates by $H\psi_E(\vec r) = E\psi_E(\vec r)$. For each potential $V(\vec r)$, given an energy $E\in \mathbb{R}$ we can define the LDOS
\begin{equation}
\rho_E(\vec r) =\sum_{E'} \mu_{|E-E'|}|\psi_{E'}(\vec r)|^2
\end{equation}
where $\mu_X:\mathbb{R}\to [0,1]$ is a weighting function centered at $X=0$ that defines an energy window. We work in units where $a=t=\hbar=1$. \par 

Given the inputs $\{V(\vec r),E,\mu_X\}$, the LDOS is well-defined and straightforward to calculate numerically. We refer to this as the ``forward problem". In practice, one may draw the inputs from distributions denoted as $\mathcal{D}_V,\mathcal{D}_E,$ and $\mathcal{D}_\mu$, respectively, whose product distribution we denote as $\mathcal{D}$. In our results, $E$ and $\mu$ are chosen deterministically. The focus of this paper is on the \textit{inverse} problem: to recover the potential $V(\vec r)$ from the LDOS $\rho_E(\vec r)$ for a given distribution $\mathcal{D}$. \par 
Note that we only require a ``slice" of the LDOS at a particular energy $E$, not all energies. 
However, this is not an essential restriction either physically or computationally. While solving the inverse problem with this restriction is strictly more difficult, relaxing this to allow $\rho_E(\vec r)$ over all energies would be interesting as well.\par 
To see that this problem may be difficult, we first point out that the problem is ill-posed in general. That is, distinct 
Hamiltonians can produce the same 
LDOS at a given energy. As a simple example, consider two distinct Hamiltonians and choose a narrow energy window that lies inside both of their spectral gaps (\textit{i.e.}, the complement of their spectrum). For both Hamiltonians, $\rho_E(\vec r) = 0$ for this energy window. However, this does not rule out the possibility of analytical solutions in less general settings of this problem or approximate solutions, especially for wider energy windows.

\paragraph*{Approach and results.} In this work, we investigate whether approximate solutions to the inverse problem are possible 
using supervised machine learning. Intuition suggests that the LDOS should reflect \textit{some} features of the potential, and approximate recovery may be possible. Moreover, we expect that the LDOS at a given position depends most strongly on the potential nearby.
Therefore, it is reasonable that a convolutional neural network (CNN) based learning approach may be suitable.
\par
\begin{figure}
    \centering
\includegraphics[width=\linewidth]{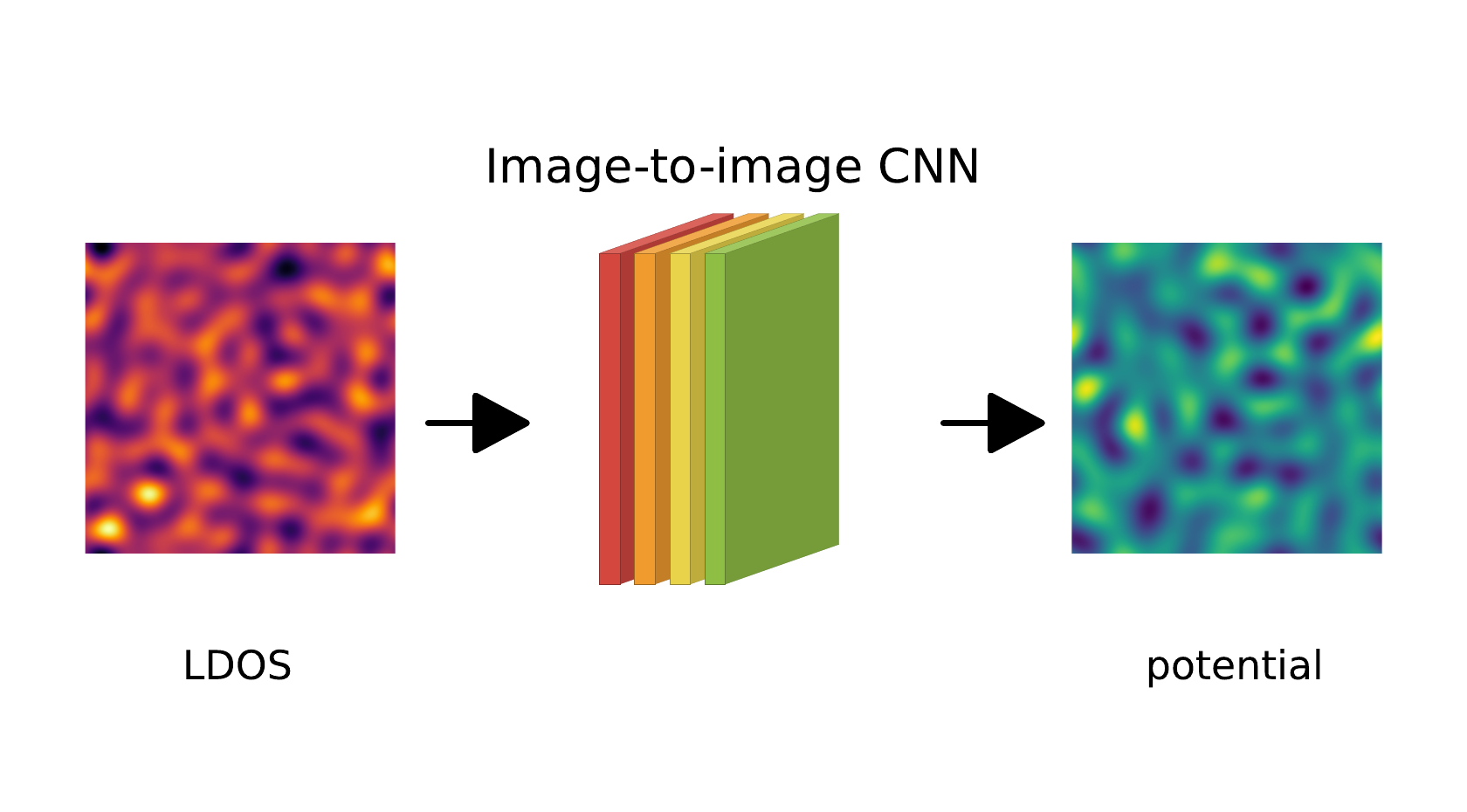}
\caption{\textbf{Potential from LDOS. } A supervised learning approach using convolutional neural networks may reconstruct the single-particle potential landscape from the local density of states. Reconstruction is possible with high quality using only modest computational resources.}
    \label{fig:schematic}
\end{figure}

 
Across a range of 1D and 2D tight-binding models with random disorder, we demonstrate that a CNN (Fig.~\ref{fig:schematic}) can invert the LDOS to reconstruct the underlying potential with high fidelity, requiring only modest computational resources ($< 1$ GPU hour). Because our CNN maps images to images, and not images to labels, we refer to our models as image-to-image CNNs. Our results suggest that Hamiltonian reconstruction from the LDOS via supervised learning is both feasible and tractable. 
Moreover, we find appreciable generalization to out-of-distribution conditions and develop models that are robust to moderate noise, suggesting that our approach could readily be applied even in settings where the underlying distribution of Hamiltonians is not precisely known.  
\par

\par

\par 
\paragraph*{Outline.}An outline of this work is as follows. In Section \ref{sec:approach}, we introduce methods for solving the inverse problem, including baselines and the image-to-image CNN. In Section \ref{sec:results}, we apply image-to-image CNNs to the inverse problem for 1D and 2D tight-binding models and study robustness to noisy inputs and distribution shifts. Finally, in Section \ref{sec:discussion}, we discuss possible experimental applications. \par 


\section{Approach}
\label{sec:approach}
Our strategy will be to train on instances of the forward problem $(V(\vec r),\rho_{E}(\vec r))$, where $V(\vec r)$ is drawn from $\mathcal{D}_V$. We normalize $V(\vec r)$ and work instead with
\begin{equation}
    \tilde V = (V-\text{mean}(V))/\text{std}(V)
\end{equation}
where std is standard deviation. We denote the prediction by $\tilde V^{\text{pred.}}$.  We do the same to the LDOS, defining 
\begin{equation}
\tilde \rho_E = (\rho_E - \text{mean}(\rho_E))/\text{std}(\rho_E). 
\end{equation}
We wish to minimize the mean-squared error (MSE) loss, which takes the form
\begin{equation}
    \mathcal{L} = \frac{1}{N_{\text{samples}}} \sum_{\text{samples}}\ell(\tilde V,\tilde V^{\text{pred.}})
\end{equation}
where the squared error (SE) loss is defined as 
\begin{equation}
        \ell(\tilde V,\tilde V^{\text{pred.}}) = \frac{1}{L^d} \sum_{\vec r} \left(\tilde V(\vec r) - \tilde V^{\text{pred.}}(\vec r)\right)^2
\end{equation}
in $d$ dimensions. We note that $\mathcal{L}=1$ on average for a predictor that predicts $\tilde V^{\text{pred.}} = 0$. We assume the lattice is square with periodic boundary conditions and side length $L$. Our focus will be on a supervised learning approach using CNNs. To give a sense of the scale of $\mathcal{L}$ and compare this approach with others, we evaluate a few baseline approaches as well. The approaches are:\par 
\paragraph{Single-parameter fit.}We study the ans\"atz $\tilde V^{\text{pred.}}(\vec r) = \alpha \tilde \rho_E^{\text{n}}(\vec r)$, where $\alpha$ is a single parameter fit to the data. This is motivated by the fact that the LDOS is often extremal where $V(\vec r)$ is extremal, at least near the edges of the spectrum. \par 
\paragraph{$k$-nearest-neighbors.}The nearest-neighbors approach is to return the $\tilde V$ corresponding to the LDOS in the training set that is closest (in Euclidean norm) to the query $\tilde \rho_E$. $k$-nearest-neighbors ($k$-nn) returns the weighted average of the potentials corresponding to the $k$ closest occurrences in the training set, weighted by inverse distance. In all cases, we also augment the data set to include all translated copies of the data (translation-augmented $k$-nearest neighbors), which we expect improves performance. 
\paragraph{Image-to-image CNN.}We use image-to-image convolutional neural networks, which in our context represent mappings from $\tilde \rho_E^{\text{n}}(\vec r) \mapsto \tilde V(\vec r)$. We emphasize that image-to-image CNNs differ from the standard CNNs that are used to predict class labels in image classification; the latter are models that map from an image to a vector of probabilities (where each entry indicates the predicted probability for a given class), rather than to another image.
Our image-to-image CNNs consist of convolutional layers with $N_c$ channels and ReLU activations (and for the 2D case, we also include residual connections); it does not use pooling or fully connected layers. 
Further details of the image-to-image CNN architectures are provided in Appendix~\ref{sec:ml}. We note that in the broader ML literature, CNNs have been used to map images to images in various contexts, such as denoising and deblurring~\cite{zhang2017beyond,nah2017deep}.

\begin{table}[!t]
  \centering
  \caption{$\mathcal{L}$ averaged over the test set for 1D and 2D studies. Single-parameter fit (1-par.) and translation-augmented $k$-nearest neighbors ($k$-nn) are compared with the image-to-image CNN. }
  \label{tab:baseline}
  \setlength{\tabcolsep}{8pt}
  \renewcommand{\arraystretch}{1.15}
  \begin{tabular}{@{}lcccc@{}}
    \toprule
    Dim. &
    1-par. &
    $1$-nn &
    $10$-nn  &
    CNN \\
    \midrule
    1D &  0.995 & 1.328 & 0.707  & \textbf{0.016} \\
    2D &  0.128 & 1.362 &0.623  &\textbf{0.005} \\
    \bottomrule
  \end{tabular}
\end{table}

\section{Results}
\label{sec:results}
\subsection{One dimensional case}
We begin in one dimension and choose $V(x)$ as a Gaussian random field with zero mean and $\langle V(x)V(x')\rangle_{\mathcal{D}_V} = V_0^2 \exp(-(x-x')^2/2\xi^2)$ (a version of the the Anderson model~\cite{Anderson1958Mar}), where $\xi$ is a spatial correlation length. We take $V_0=0.5, E=-1.5,\xi = 3$, assume noise is absent, and take an energy window $\mu_X =\mathbbm{1}(|X|<\delta E/2)$ with $\delta E = 0.25$. 

\par 
\begin{figure}
    \centering
\includegraphics[width=\linewidth]{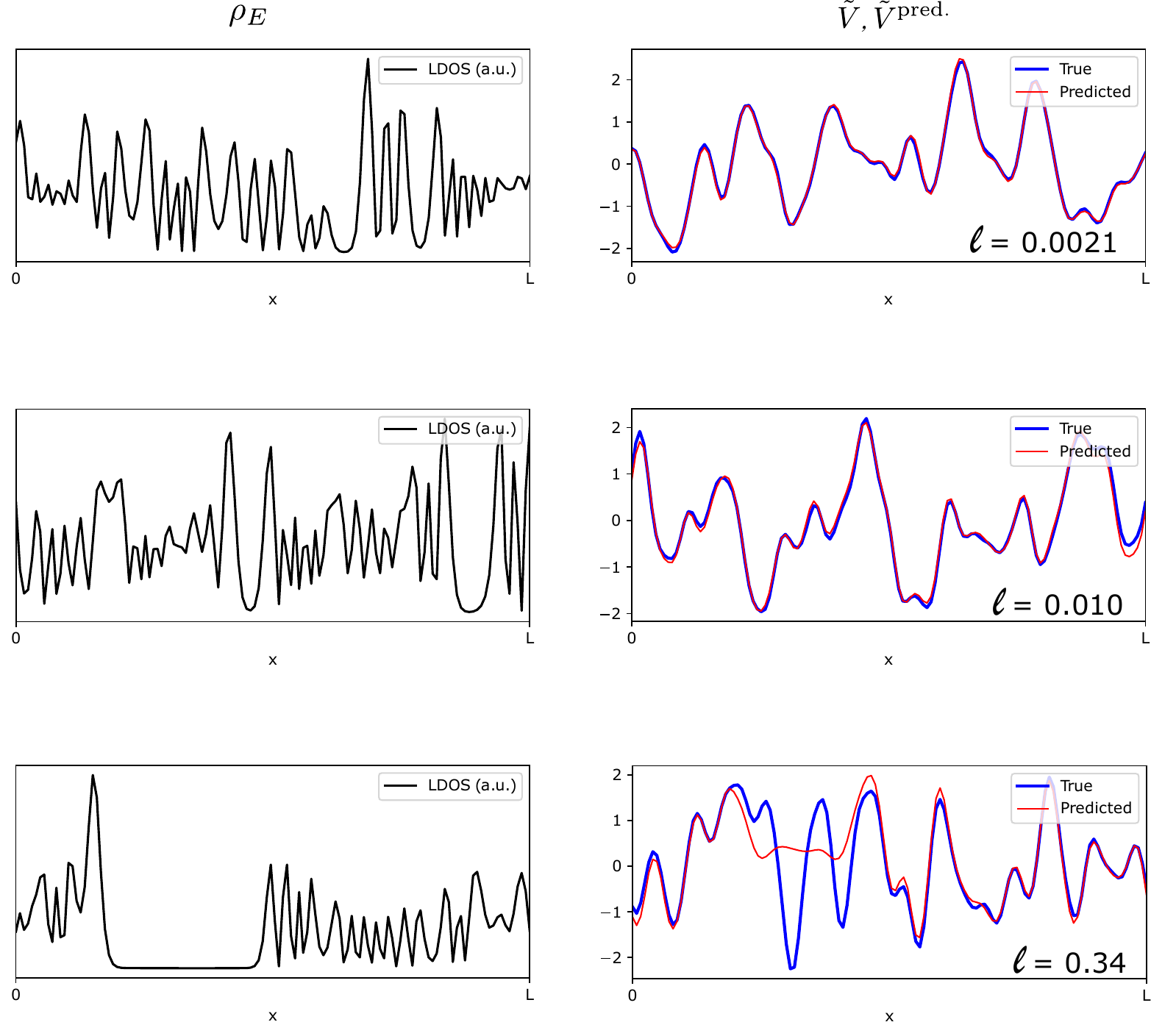}    \caption{\textbf{1D case.} Examples for the 1D case (NN-1D), which takes as input the local density of states $\rho_E(x)$ of electrons in a 1D chain with correlated disorder (right) and predicts the potential landscape (right). From top to bottom: best case, median case, and worst case in the test set, with indicated squared error losses $\ell = \ell(\tilde V, \tilde V^{\text{pred.}})$. The average MSE loss was $\mathcal{L} = 0.016$. }
    \label{fig1}
\end{figure}
\begin{figure}
    \centering
\includegraphics[width=\linewidth]{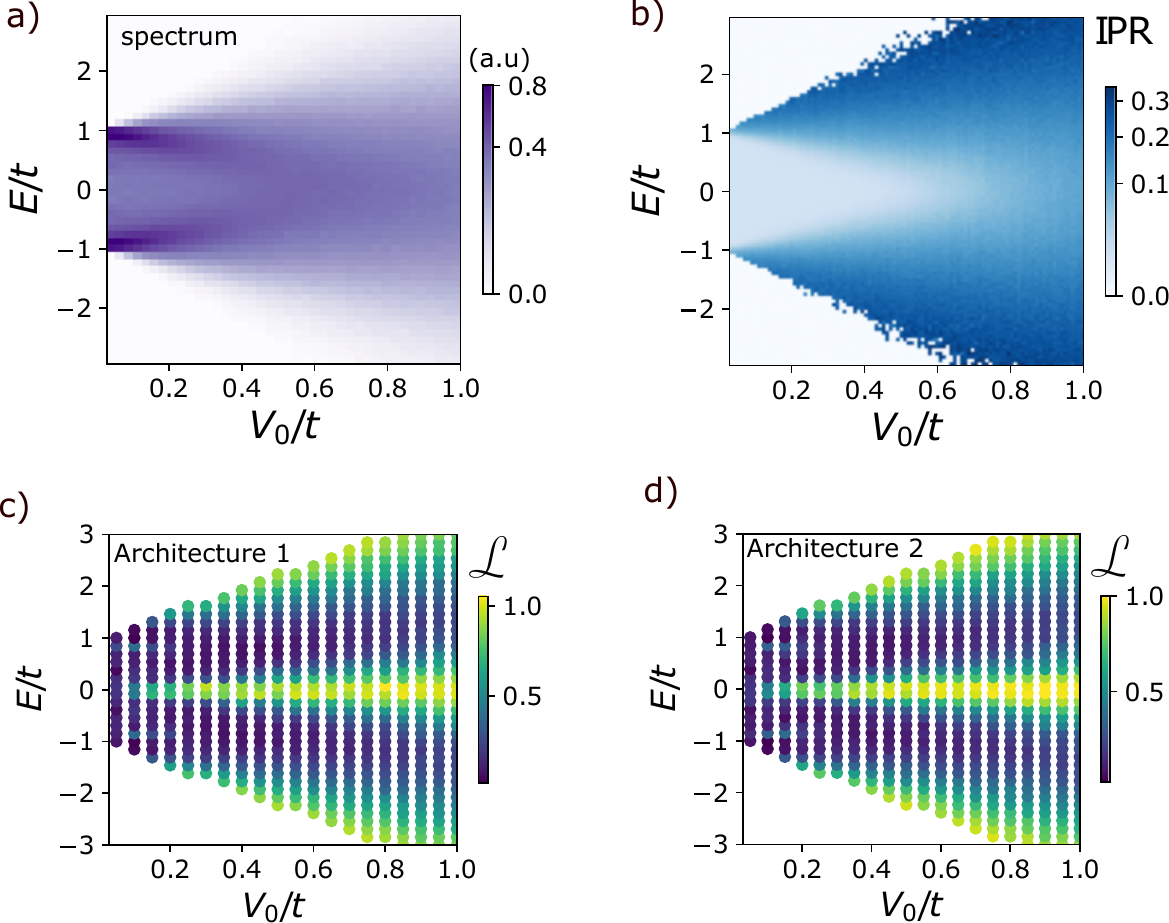}
\caption{\textbf{Localization and learnability. }(a) Spectrum of 1D Anderson model as a function of the potential strength $V_0$ ($\xi=3$). (b) Average inverse participation ratio (IPR), given by $\sum_{\vec r}|\psi_{\vec r}|^4$. (c,d) Final MSE loss for two simple architectures (see Appendix~\ref{sec:ml}). Smaller $\mathcal{L}$ indicates greater ``learnability".  }
    \label{fig:1dheatmap}
\end{figure}
 For system size $L=128$, we generated $12000$ data points, tuned hyperparameters, and trained for 50 epochs. We achieve $\mathcal{L} \approx 0.016$ on the test set, greatly exceeding the performance of baselines (Tab. \ref{tab:baseline}). We refer to this trained model as \textbf{NN-1D}. We show the results of NN-1D on four test LDOS profiles in Fig.~\ref{fig1}. 
 
 In the absence of a potential, the wavefunctions are plane waves with wavevector $k$ at energy $E$ as determined by the dispersion $E(k) = -t\cos(k)$. As the potential is turned on, the wavefunctions Anderson localize with a localization length that increases with increasing $V_0$ and $|E|$. We illustrate the spectrum and localization properties in Fig.~\ref{fig:1dheatmap}.\par

The quality of the reconstruction depends on the choices of certain parameters. To observe this, we choose two simple architecture, vary $V_0$ and $E$, and plot final MSE losses in Fig.~\ref{fig:1dheatmap}c,d. For this analysis, the main goal was to capture how each parameter generally affects the reconstruction, and we did not optimize for low losses or tune any hyperparameters. We can observe that the model's worst performance occurs near the middle and edges of the spectrum at any fixed $V_0$. This indicates that the performance is not strictly correlated with either the spectral density (or sparsity) or the localization length of the eigenstates, although we may expect both of these quantities to play a role.
\par 
\subsection{Two dimensional case \label{subsection_two_dimensional_case} }

Next, we will show that the high reconstruction quality of the Hamiltonian extends to two dimensions. As described previously, we choose $V(\vec r)$ to be a Gaussian random field with zero mean and $\langle V(\vec r)V(\vec r')\rangle_{\mathcal{D}_V} = V_0^2\exp(-(\vec r-\vec r')^2/2\xi^2)$ with $V_0 = 0.5,\xi = 3$. We train on $L\times L$ systems with $L=64$; other differences with the 1D case are that we generated $1200$ data points, chose $E = -1.0$ and included residual (skip) connections for every two convolutional layers. \par
\begin{figure}
    \centering
\includegraphics[width=\linewidth]{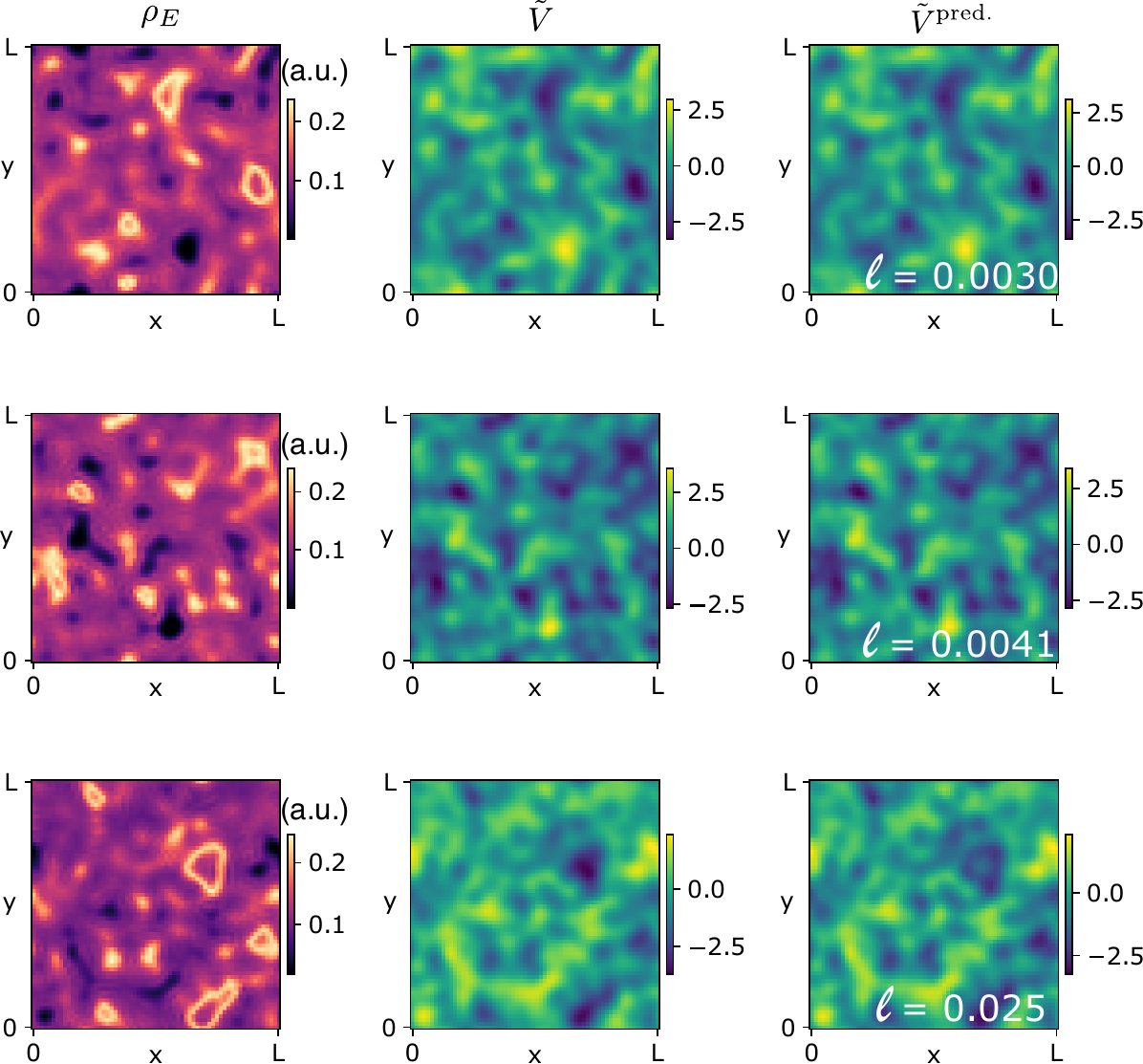}
    \caption{\textbf{2D case. } Examples for the 2D case (NN-2D). From left to right: the LDOS, true potentials, and predicted potentials. From top to bottom: best case, median case, and worst case in the test set, with indicated square error losses $\ell = \ell(\tilde V,\tilde V^{\text{pred.}})$. The average MSE loss was $\mathcal{L} \approx 4.6 \cdot 10^{-3}$.}\label{fig:2d}
\end{figure}

 
After hyperparameter tuning, the trained model achieves $\mathcal{L} \approx 4.6 \times 10^{-3}$ on the test set, again greatly exceeding the performance of baselines (Tab. \ref{tab:baseline}). (As a caveat, we expect the nearest-neighbors baselines will improve as the training set size increases, so the comparison against baselines is only meaningful for fixed training set size). We refer to this trained model as \textbf{NN-2D}. Details of the architecture are described in Appendix \ref{sec:ml}. We show the results of NN-2D on three test LDOS profiles in Fig.~\ref{fig:2d}, noting that the predicted and true potentials are almost indistinguishable. 
\par 
Some features of the LDOS are worth noting in Fig.~\ref{fig:2d}. First, due to the low energy $E=-1.0$, there are LDOS vacancies (black regions) corresponding to where $V(\vec r)$ is large. Second, there are rings of charge near potential wells, which act as effective quantum dots or harmonic traps, which are commonly observed in STM. 

\subsection{Robustness}
\label{sec:robust}
Having established that a suitably trained image-to-image CNN can accurately reconstruct the potential landscape in both 1D and 2D, we now consider more challenging scenarios that may be relevant in realistic experimental scenarios. In actual experiments, noise is an inevitable factor, and the underlying system parameters (such as correlation length or energy windows) may be unknown \textit{a priori}. In this section, we study the performance of (variants of) NN-2D when the test set is adversely affected by either noise or distribution shifts. 

\paragraph*{Noise resilience.}

The first adverse effect we consider is the presence of noise in the test set. In addition to assessing the impact of adverse test noise across a range of noise levels, we also implement a training approach designed to improve robustness to test noise.
Our approach for improving robustness is based on intentionally adding synthetic noise to the LDOS samples in the training set.
The fact that it can actually be beneficial to intentionally add noise to the training set has been well-studied in the broader deep learning literature; in particular, adding training noise can be viewed as a form of regularization~\cite{an1996effects,bishop1995training,karpukhin2019training}.

\begin{figure}
    \centering    \includegraphics[width=0.6\linewidth]{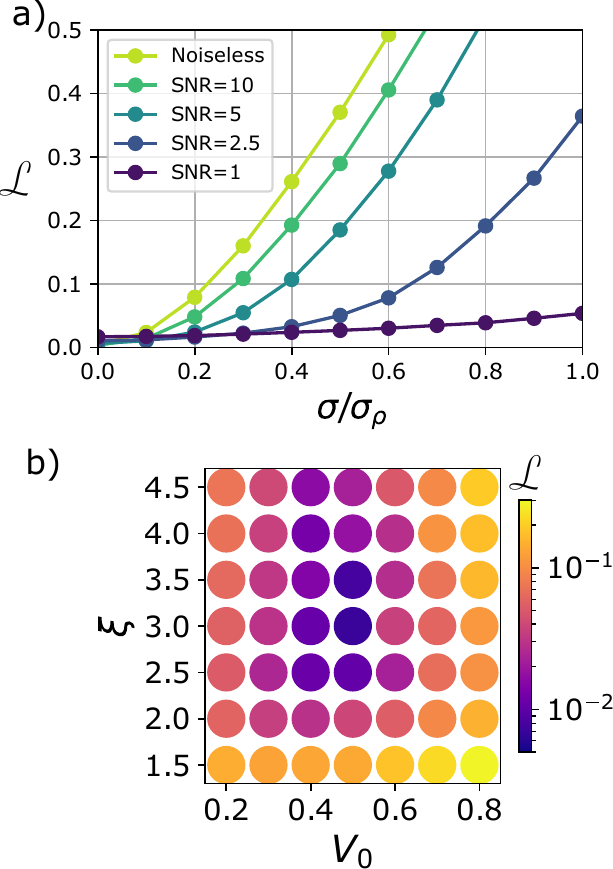}   \caption{\textbf{Robustness. } (a) Test-time performance for variants of NN-2D with varying signal-to-noise ratios (SNRs) in the training data. The noiseless case is identical to NN-2D. $\sigma/\sigma_\rho$ denotes ratio of test sample noise to average LDOS amplitude. (b) Average test MSE for NN-2D across ten disorder realizations at each amplitude $V_0$ and correlation length $\xi$, testing OOD generalization.}
    \label{fig:robust}
\end{figure}

In Figure~\ref{fig:robust}a, we present empirical results for spatially correlated test and train noise.
The yellow curve (labeled ``noiseless'') corresponds to using the same NN-2D model described in Section~\ref{subsection_two_dimensional_case} (which was trained on noiseless data), and the other curves correspond to retraining NN-2D on LDOS images corrupted by correlated Gaussian noise ($\xi_{\text{noise}} = 3$). We considered various levels of signal-to-noise ratios (SNR) when injecting training noise (note that noiseless is equivalent to SNR = $\infty$). 
The yellow curve indicates that, even without regularization, the network tolerates \(\sim\!20\%\) correlated  test noise (\(\sigma/\sigma_\rho = 0.2\)) while maintaining $\mathcal{L} <0.1$. Here $\sigma/\sigma_\rho$ is the ratio of the noise amplitude to the LDOS fluctuation amplitude. Moreover, we find that regularizing by including noisy samples during training markedly boosts robustness: the variant trained on SNR\,$=1$ noise maintains $\mathcal{L}$ below \(0.05\) out to almost a test $\sigma/\sigma_\rho$ value of unity. These results suggest that modest regularization is sufficient for reliable reconstruction under realistic experimental conditions where thermal and instrumental noise do not exceed the LDOS signal itself. Further details, including empirical results for spatially uncorrelated test noise (where we use spatially uncorrelated train noise to regularize), are presented in Appendix \ref{app:noise}.

\textit{Distribution shifts. } Realistic scenarios could involve deploying a model on test samples whose true description differs from what was assumed in training, creating distribution shifts where an otherwise successful model may fail to extrapolate. The capacity to perform well under such distribution shifts \(\mathcal{D}\to\mathcal{D}'\) is crucial if one aims to apply learned reconstructions to real systems, where parameters such as correlation length, potential strength, energy windows, or noise levels may be unknown \textit{a priori}. \par 
We assess resilience to distribution shifts in NN-2D by varying two parameters: the disorder amplitude \(V_0\) and correlation length \(\xi\). We sample \(\bigl(V_0,\xi\bigr)\) values in the neighborhood of the training value \(\bigl(0.5,3.0\bigr)\) and compute the mean-squared error on data newly generated using these shifted parameters, plotting results in Fig.~\ref{fig:robust}b. This approach indicates how far the model’s accuracy persists once the underlying distribution departs from \(\mathcal{D}\). Figure~\ref{fig:robust}b shows that NN-2D continues to achieve low errors near its training point, suggesting it is not merely learning narrowly applicable principles but can recover nearby potential landscapes as well. For instance, \(\mathcal{L} \lesssim 0.1\) for \(0.3\leq V_0\leq 0.6\) and \(2.0\leq \xi\leq 4.5\). For a visual sense of reconstruction quality at \(\mathcal{L}\sim 0.1\), see Fig.~\ref{fig:oodsupp}. 
\par 
Overall, these trends indicate that NN-2D is generalizable not only out-of-sample but also out-of-distribution. We may note that this distribution shift study did not involve using any regularization, and it is plausible that adding regularization could improve the results. 
Moreover, in our studies so far, we have restricted to the LDOS at a \textit{particular} energy $E$. In practical applications, the energy $E$ can be easily and precisely tuned (e.g. by bias voltage in STM) and the LDOS can be imaged over a \textit{range} of energies. This provides an additional dimension of information to train on that could improve the reconstruction quality of the potential. This could be useful or even essential when the noise amplitudes or distribution shifts are drastic. In sum, various strategies could bolster the model's robustness and help bridge the gap to realistic applications.\par

\section{Discussion} 
\label{sec:discussion}


\paragraph*{Applications for STM.} Scanning tunneling microscopes measure local conductance $I(\vec r)$ as a function of bias voltage $V_{\text{b}}$ with extraordinary sensitivity. In the single-particle limit, the LDOS of the sample is related to the differential conductance between the tip and sample $\rho_{E_F-eV_{\text{b}}}(\vec r) \propto dI(\vec r)/dV_{\text{b}}$. The result is a real-space image of the electronic LDOS with sub-angstrom resolutions. Machine learning has been applied to extract nontrivial information from STM images, for instance signatures of nematicity~\cite{Zhang2019Jun,Sobral2023Aug} and defects or lattice distortions~\cite{Choudhary2021Feb,Shi2022May}. \par 

In our work, we have established that for certain non-interacting lattice systems, the local potential can be efficiently reconstructed by an ML approach from the LDOS at a particular energy (as would be measured by STM). A concrete application of such a reconstruction is modeling the energy landscape in a 2D 
material in order to perform downstream calculations or simulations of other physically interesting quantities (e.g. transport coefficients, responses to magnetic or displacement field, etc.) that are in principle computable once the potential is known.
\par 
A limitation for this scenario is that often STM practitioners are interested in surfaces that are quite regular, perhaps with a low density of defects. In these instances, characterizing the energy landscape becomes essentially a problem of locating and labelling a discrete set of defects. For instance hBN may have a defect density of $1/(100\text{ nm})^2$ and three main defect types\cite{Wong2015Nov}. In these scenarios, where the space of possible physical Hamiltonians is ``low-dimensional", we expect that an ML approach cannot offer much advantage over simple regression or human inspection. \par 

Another possible application of our approach is to the study of surface reconstructions, an area where STM has historically had great success. Surface reconstruction is the possibly complicated reorganization of the surface atoms of a bulk material along a cleaved plane. It can easily lead to complex spatial patterns, for instance the famous $7\times7$ pattern on Si(111)~\cite{Binnig1983Jan} and herringbone pattern on Au(111)~\cite{Perdereau1974May}, due to the complicated (possibly aperiodic) energy landscape formed by atoms near the surface. In the noninteracting approximation, the energy landscape at some tip-distance can be captured by an effective potential $V(\vec r)$, with $\vec r$ the surface coordinate, which our work shows can be efficiently reconstructed from the measured LDOS (see also \cite{Li2022Oct}). This could be extended to inference of the detailed atomic structure in the cleaved plane. We leave this direction for future work.\par 
\paragraph*{Comments on problem setting.}  

Our analysis has so far been confined to the setting of a single–orbital, nearest-neighbor tight-binding Hamiltonian with a site-diagonal disorder potential, a choice that enables efficient data generation and transparent error metrics. In real materials, however, we may wish to include various generalizations, most importantly electron-electron interactions. In this case, reliable synthetic data generation may become more expensive. An interesting goal along these lines is training a neural network to learn the interaction strength or other details of the interaction directly from the LDOS.

It is worth highlighting that in the problem setting we considered, we only used 1,000 samples in total for training and validation for the two dimensional case.
A more comprehensive study of performance as a function of training set size may be a valuable future direction, as it could help illuminate the limits of how little data is truly required.
Moreover, the fact that we get quite good performance with relatively little data may be an encouraging early sign that our approach may still work in situations where it is difficult to collect large amounts of training data, such as if the data were to come from expensive calculations for systems with electron-electron interactions or perhaps from experimental observation.

\paragraph*{Outlook.}

In this work, we posed a new inverse problem -- namely, whether one can reconstruct the Hamiltonian from the LDOS.
\textit{A priori}, it was not obvious whether reasonably good approximate solutions to this problem should exist at all.
Our empirical results demonstrate that in certain settings and for a given distribution, quite good approximate solutions are in fact possible.
We obtained these empirical results using machine learning, but we do not rule out the possibility of other approaches based on traditional numerical techniques or even analytics -- indeed, pursuing such approaches may be an interesting direction for future work.
More broadly, we hope our results will motivate further study of this inverse problem and provide new perspectives on Hamiltonian learning.

\bibliography{bib}

\appendix
\onecolumngrid

\begin{table}[b!]
    \caption{Summary of notation.}
    \label{tab:notation}
    \centering
    \begin{tabular}{ll}
    \toprule
    \textbf{Notation} & \textbf{Meaning} \\
    \midrule
    $V(\vec r)$ & potential\\
    $[\rho_E^{\text{n}}(\vec r)]\,\, \rho_E(\vec r)$ & [noised] LDOS\\
    $V_0$ & potential standard deviation\\
    $\xi$ & correlation length\\
    $L$ & system size\\
    $\delta E$ & energy window half-width\\
    $\eta_{\vec{r}}$ & noise  \\
    $\mathcal{D}$ & distribution of data\\
    $N$ & \# of samples\\
    $N_{\text{ep.}}$ & \# of epochs\\
    $b$ & batch size \\
    $k$ & convolutional kernel size \\
    $N_c$ & \# of channels \\
    $N_L$ & \# of layers\\
    $\mathrm{lr}$ & learning rate \\
    \bottomrule
    \end{tabular}\label{table}
\end{table}

\section{Data generation, ML implementation, and baselines}
\label{sec:ml}
Numerical experiments were implemented in PyTorch 2.0 and conducted on an Nvidia A100 GPU (neural networks) or CPU (baselines). We trained image-to-image CNNs with batch size $b$, convolutional kernel size $k$, learning rate lr, $N_c$ channels, $N_L$ layers, and ReLU activations. Notation is summarized in Table \ref{table}. 

\subsection{Synthetic data generation}
We generated $N $ synthetic data points with a $2/3-1/6-1/6$ train-validation-test split. To produce the random potentials $V(\vec{r})$ discussed in the main text, we generate a Gaussian random field in both one and two dimensions by first creating white noise in reciprocal space and then multiplying by \(\sqrt{\text{psd}(\vec{k})}\), where the power spectral density (psd) is of the form $\text{psd}(\vec{k}) \,\propto\, \exp\Bigl(-\tfrac{|\vec{k}|^2\,\xi^2}{2}\Bigr).$
An inverse discrete Fourier transform returns $V(\vec{r})$ in real space. We used periodic boundary conditions. Disorder potentials were chosen to be Gaussian with amplitude $V_0 = 0.5$ and correlation length $\xi = 3$ (recall $a=1$ is the lattice constant). 
\par 
We defined the local density of states (LDOS) near energy $E$ as 
\begin{equation}
\rho_E(\vec{r}) \;=\; \sum_{E'} \,\mathbbm{1}[|E - E'| < \tfrac{\delta E}{2}] \;|\psi_{E'}(\vec{r})|^2.
\end{equation}
For the case with noise, we added noise in real space by $
   \rho_E^{\text{n}}(\vec{r})
   \;=\;
   \rho_E(\vec{r})
   \;+\;
   \eta_{\vec{r}}$ 
where $\eta_{\vec{r}}$ was generated in the same way as $V(\vec r)$. We applied a global normalization scheme to both the LDOS and the potential:
\begin{equation}
   \tilde{\rho}_E(\vec{r})
   \;=\;
   \frac{\rho_E(\vec{r}) - \langle \rho_E\rangle}{\sigma_{\rho}},
   \quad\quad
   \tilde{V}(\vec{r})
   \;=\;
   \frac{V(\vec{r}) - \langle V\rangle}{\sigma_{V}},
\end{equation}
where $\langle \cdot \rangle$ and $\sigma$ denote the mean and standard deviation taken over the entire training set.  The image-to-image CNNs thus learn a dimensionless mapping from $\tilde{\rho}_E \mapsto \tilde{V}$.  We did not attempt sample-wise normalization.

\subsection{Neural networks}

\paragraph*{NN-1D} For NN-1D, we employed a 1D CNN consisting of $N_L$ hidden layers with constant channel width. The network takes 1D input signals of length $L$ and applies a sequence of 1D convolutions with circular padding to preserve periodicity using the standard PyTorch module \texttt{Conv1d}. Each hidden layer contains $N_c$ channels, with kernel size $k$ applied across all layers, and padding $\lfloor k/2\rfloor$. The architecture follows the pattern: $1 \to N_c \to N_c \to \ldots \to N_c \to 1$, where the first layer expands from single-channel input to $N_c$ feature channels, $N_L-1$ intermediate layers maintain $N_c$ channels with ReLU activations, and a final layer projects back to single-channel output. $N_c, N_L,k$ and lr were tuned using Optuna. Using the \texttt{suggest\_int} and \texttt{suggest\_int} features, these were tuned in the vicinities of $N_c \in [16,256], N_L \in [4,16], k \in [3,9]$ and lr $\in [10^{-4},10^{-2}]$. Optuna training was done with the train set, validated on the validation set, and limited to 20 epochs and 30 trials. The best architecture was chosen using the validation loss and resulted in $N_c = 112, N_L = 6, k = 5, \text{lr} \approx 1.03 \times 10^{-3}$. Finally, we trained NN-1D on the training set for $N_{\text{ep.}} = 50$ epochs using batch sizes $b=256$, energy scales $E = -1.5,\ \delta E = 0.25$, and system size $L=128$. The total number of data points was $N = 15000$. 

\par 

\paragraph*{NN-2D} For NN-2D employed a 2D residual neural network (equivalently, a CNN with skip connections) consisting of $N_L$ convolutional layers. The network takes 2D input signals of size $L \times L$ and applies a sequence of $N_B$ residual blocks with circular padding to preserve periodic boundary conditions. The architecture consists of an initial projection layer that expands from single-channel input to $N_c$ base channels, followed by $N_B$ residual blocks that maintain constant channel width, and a final projection layer back to single-channel output. Each residual block contains two $k \times k$ convolutional layers with batch normalization and ReLU activations, connected by a skip connection. The total layer count relationship is $N_L = 2N_B + 2$. We again tuned hyperparameters using Optuna in the vicinities of $N_c \in [16,128], N_B \in [2,12], k \in [3,9]$ and lr $\in[10^{-4},10^{-2}]$. Optuna training was done with the train set, validated on the validation set, and limited to 7 epochs and 50 trials. The best architecture was chosen using the validation loss and resulted in $N_c = 32, N_B = 3, k = 3,$ lr $\approx 7.21\times 10^{-4}$. Finally, we trained NN-2D on the training set for $N_{\text{ep.}} = 50$ epochs using batch sizes $b=16$, energy scales $E=-1.0, \delta E = 0.25$, and system size $L=64$. The total number of data points was $N=1200$. 

\par

\paragraph*{Other details} We used the Adam optimizer to minimize the mean-squared error (MSE) loss. We reduced the learning rate upon a validation loss plateau using the standard \texttt{ReduceLROnPlateau} with factor $=0.1$ and patience $=3$. Although our experiments fix $L=128$ for 1D and $L=64$ for 2D, we briefly tested smaller sizes in pilot runs.  The same training pipeline completes much faster but no qualitatively different behaviors were observed, with the network still learning with comparable MSE loss. \par 


\subsection{Baselines}
We compared our neural network approach against two baseline methods for the inverse mapping from normalized local density of states $\tilde{\rho}_E(\vec{r})$ to normalized disorder potential $\tilde{V}(\vec{r})$.

The single-parameter linear fit was the first baseline. It assumes a direct proportionality between (mean-subtracted) LDOS and potential:
$
\tilde{V}^{\text{pred}}(\vec{r}) = \alpha \tilde{\rho}_E(\vec{r})
$
where $\alpha$ is determined by least-squares fitting on the training data:
\begin{equation}
\alpha^* = \frac{\sum_{i=1}^{N_{\text{train}}} \sum_{\vec{r}} \tilde{\rho}_E^{(i)}(\vec{r}) \tilde{V}^{(i)}(\vec{r})}{\sum_{i=1}^{N_{\text{train}}} \sum_{\vec{r}} [\tilde{\rho}_E^{(i)}(\vec{r})]^2}
\end{equation}
This ansatz is motivated by the anti-correlation between LDOS and potential at low energies, where states tend to localize in regions of lower potential. It seems to work well in 2D but not 1D. \par 

\par 
Translation-augmented $k$-nearest neighbors was the second baseline. It searches for training samples with LDOS patterns most similar to the query and returns a weighted average of their corresponding potentials. For a query LDOS $\tilde{\rho}_E^{\text{query}}(\vec{r})$, we: (1) Generate all spatial translations of each training pair $(\tilde{\rho}_E^{(i)}, \tilde{V}^{(i)})$ on a coarse grid with step size $s=1$ for 1D and $s=8$ for 2D (yielding $64$ augmented copies per original sample for $64 \times 64$ lattices), (2) Calculate cosine distances between the flattened query LDOS and all augmented training LDOS patterns:
\begin{equation}
d_j = 1- \frac{\langle x,x_j\rangle}{\|x\|_2\|x_j\|_2}
\end{equation}
where $x = \text{vec}(\tilde{\rho}_E^{\text{query}}), x_j =\text{vec}(\tilde{\rho}_E^{(j)})$ and $\text{vec}(\cdot)$ denotes vectorization of the 2D field, (3) Return the distance-weighted average of potentials from the $k$ nearest neighbors:
\begin{equation}
\tilde{V}^{\text{pred}}(\vec{r}) = \frac{\sum_{j \in \mathcal{N}_k} w_j \tilde{V}^{(j)}(\vec{r})}{\sum_{j \in \mathcal{N}_k} w_j}
\end{equation}
where $\mathcal{N}_k$ denotes the set of $k$ nearest neighbors and $w_j = 1/d_j $. Cosine distances were favored over Euclidean distances because they performed better in our case. The translation augmentation exploits the translational symmetry of the Anderson model with periodic boundary conditions, effectively increasing the training set size by a factor of 128 (for 1d) or 64 (for 2d) while preserving the underlying physics.\par 

\par

In both cases, the training data was the union of the train and validation sets used for the neural networks. Since the baselines do not need separate validation sets, this ensures that the same amount of data was exploited by all of the models in the model-building process. In particular, this was $12000$ points for the 1d case and $1000$ points for the 2d case. 

\subsection{Figure \ref{fig:1dheatmap} and out-of-distribution study}

For Fig. \ref{fig:1dheatmap}c-d, we performed a ``learnability" study of the 1D Anderson model. We chose system size $L = 64$, correlation length $\xi = 3$, and disorder amplitudes $V_0 /t \in[0.05, 1.0]$. For each disorder amplitude $V_0$, we generated 500 random disorder realizations, diagonalized each Hamiltonian, and stored results for reuse across different energy windows. We employed a 4/5-1/5 train-test split. For each target energy window, we computed the LDOS across an energy window $\delta E /t = 0.25$ to create training pairs. We focused on a physically motivated region of paramater space: $E \in [-3,3]$ with $|E|< |2.5 V_0 + 1|$. Outside of these bounds, the spectrum is extremely sparse, leading to poor training. We trained two image-to-image CNNs, Architectures 1 and 2, which differ from NN-1D only in hyperparameters. The hyperparameters were $b=32, k=5,N_c = 64, N_L = 5$ and $b=16, k=7, N_c = 32, N_L = 3$, respectively. We trained with Adam for 10 epochs with lr = $10^{-3}$ in both cases, and plotted the final MSE losses averaged over the test set.

\par
\begin{figure}
    \centering    \includegraphics[width=0.5\linewidth]{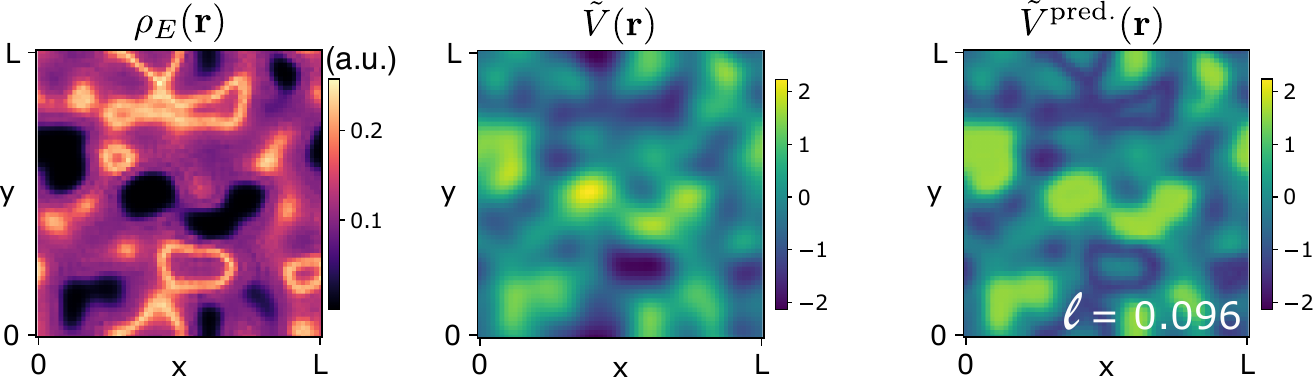}
    \caption{\textbf{Out-of-distribution sample. }Evaluation of NN-2D on a sample with $V_0 = 0.8, \xi = 4.5$ with $\ell =0.096$. Two common failure modes are imprints of charging rings near potential minima and plateaus near potential maxima.}
    \label{fig:oodsupp}
\end{figure}

 
For the out-of-distribution study (Figure \ref{fig:robust}b) we generated 10 new pairs of potential and LDOS for each grid point, keeping all parameters the same as the NN-2D study except for $V_0$ and $\xi$. We evaluated NN-2D on these points and averaged the MSE loss to produce the figure. An example of the performance of NN-2D out-of-distribution is in Fig.~\ref{fig:oodsupp}.

\section{Addition of noise}
\label{app:noise}

\label{sec:robustness}
\begin{figure}
    \centering
\includegraphics[width=\linewidth]{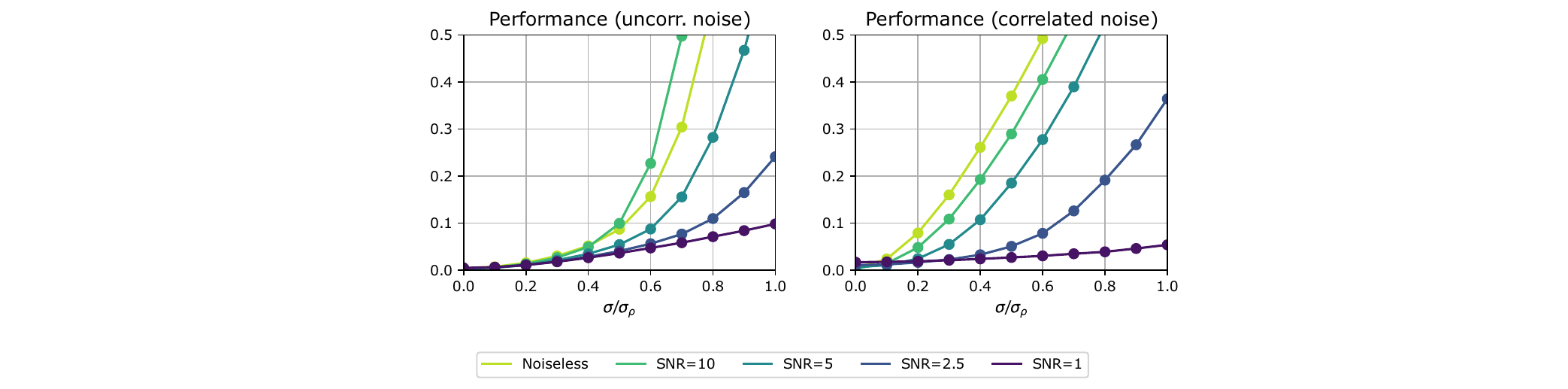}
    \caption{\textbf{Noise robustness. }Test performance for variants of NN-2D with varying signal-to-noise ratios (SNRs) in the training data. The noiseless case is identical to NN-2D. $\sigma/\sigma_\rho$ denotes the ratio of test sample noise to average LDOS amplitude. Training and test noise is uncorrelated (left) and correlated with $\xi_\eta = 3$ (right, identical to Fig. \ref{fig:robust}a). }
    \label{fig:noise-sweep}
\end{figure}
\begin{figure}
    \centering
\includegraphics[width=0.75\linewidth]{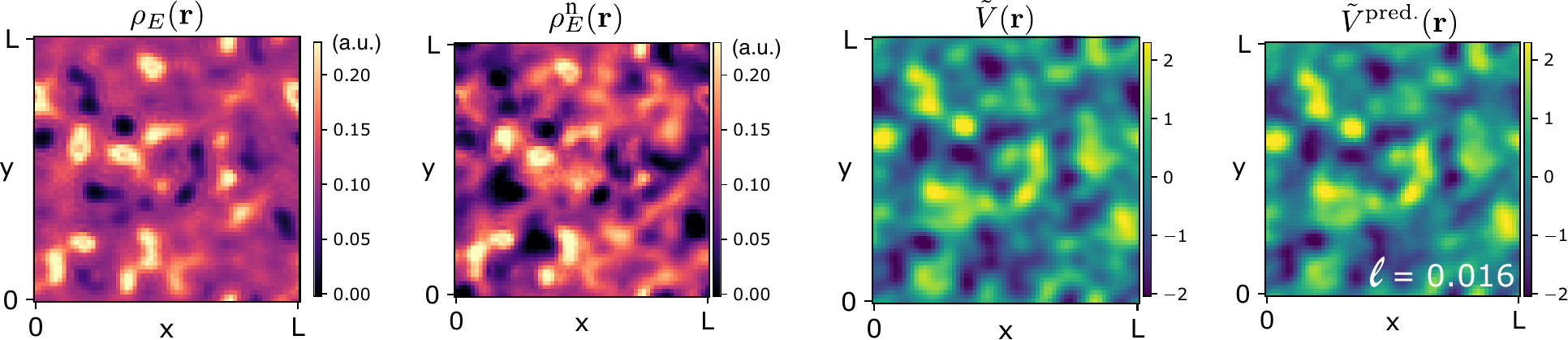}
    \caption{\textbf{A noisy sample. } Evaluation for the model trained on correlated noise with SNR $=1$ (c.f. Fig.~\ref{fig:noise-sweep}) on a noised LDOS with $\sigma/\sigma_\rho =1$. (a) Noiseless and (b) noised LDOS. (c) True and (d) predicted $V(\vec r)$.}
    \label{fig:noise-example}
\end{figure}

A potential challenge in deploying an LDOS $\to$ Hamiltonian solver in practical scenarios is the presence of noise in the test samples. One strategy to improve performance in this case is to add noise to the training set, which can be viewed as a form of regularization~\cite{an1996effects,bishop1995training,karpukhin2019training}. To this end, we considered both training-set noise (which can potentially have a beneficial regularizing effect) and
test-set noise (which is adverse). We proceed by studying two representative scenarios for the noise $\eta_{\vec r}$ added to the LDOS: (1) uncorrelated Gaussian noise and (2) correlated Gaussian noise with correlation length $\xi_{\eta}$. 

We train on noised LDOS data $\rho_E^{\text{n}}(\vec r) = \rho_E(\vec r) + \eta_{\vec r}$, where $\eta_{\vec r}$ is chosen sample-by-sample. For each sample, $\bar\sigma$ is first chosen uniformly in $[0,\sigma_\eta]$ and $\eta_{\vec r}$ is subsequently chosen from $\mathcal{N}(0,\bar\sigma^2)$ for each $\vec r$. For case (2), we also feed $\eta_{\vec r}$ through a low-pass filter with correlation length $\xi_{\eta}$. We label the new models by the signal-to-noise ratio
\begin{equation}
    \text{SNR} = \sigma_\rho / \sigma_\eta
\end{equation}
of the training data, where $\sigma_\rho = \sqrt{\langle \rho_E(\vec r)^2\rangle - \langle \rho_E(\vec r)\rangle^2}$. For simplicity, we use the tuned hyperparameters we obtained when creating NN-2D in the un-noised scenario and we further train on noised versions of the train set and test on noised versions of the test set. We plot performance in Fig.~\ref{fig:noise-sweep} across five models: the noiseless baseline model (SNR $=\infty$) and models with SNR $=10,5,2.5,$ and $1$. We choose a correlation length $\xi_{\eta} = 3$ for the correlated case.\par 
After training, we next evaluate each variant on test sets with artificially added noise of amplitude $\sigma$. Each data point in Fig.~\ref{fig:noise-sweep}b and Fig.~\ref{fig:noise-sweep}d represents an average MSE loss across 200 test samples. We find that noise robustness is good for small amounts of test noise and is improved significantly by adding training noise (with small SNR, i.e. large training noise), as discussed in the main text. For the most strongly regularized model we considered (SNR$=1$), the MSE loss remains $\lesssim 0.1$ even when the noise reaches $100\%$ of the LDOS amplitude. 
A test sample for the SNR$=1$ model with correlated noise (with $\sigma=\sigma_\rho$) is shown in Fig.~\ref{fig:noise-example}, along with the the corresponding true and predicted potentials. 

\par 

Our results indicate that our basic approach is fairly robust against moderate test noise
levels, especially if regularization (achieved via adding training noise in our work) is applied. That said, real STM measurements can exhibit additional complications
like sample-tip distance fluctuations, drift-induced distortions, and temperature effects. 
In practical settings, a modest level of noise in STM images is thus
unlikely to fully obscure the local potential, provided sufficient training noise is injected, although large noise or severe
drifts may demand more careful modeling. \par

\end{document}